\documentclass[sigconf]{acmart}
\pdfoutput=1

\usepackage{wrapfig}
\usepackage{dblfloatfix}
\usepackage{hyperref}
\usepackage{balance}
\usepackage{colortbl}

\usepackage{cleveref}
\usepackage[shortlabels]{enumitem}

\usepackage{pifont}

\newcommand{\cmark}{\ding{51}}%
\newcommand{\xmark}{\ding{55}}%

\newcommand{\bi}{\begin{itemize}[leftmargin=0.4cm]}
	\newcommand{\ei}{\end{itemize}}
\newcommand{\be}{\begin{enumerate}[leftmargin=0.4cm]}
	\newcommand{\ee}{\end{enumerate}}

\makeatletter
\let\th@plain\relax
\makeatother

\definecolor{Gray}{gray}{0.85}
\usepackage{tikz}
\usepackage{framed}
\usepackage[framed]{ntheorem}
\usetikzlibrary{shadows}

\theoremclass{Lesson}
\theoremstyle{break}

\tikzstyle{thmbox} = [rectangle, rounded corners, draw=black,
fill=Gray!40]

\newshadedtheorem{lesson}{Result}

\newcommand{\tion}[1]{{Section }\ref{sect:#1}}

\usepackage{listings}
\definecolor{MyDarkBlue}{rgb}{0,0.08,0.45} 
\copyrightyear{2018} 
\acmYear{2018} 
\setcopyright{acmcopyright}
\acmConference[ICSE '18]{ICSE '18: 40th International Conference on Software Engineering }{May 27-June 3, 2018}{Gothenburg, Sweden}
\acmBooktitle{ICSE '18: ICSE '18: 40th International Conference on Software Engineering , May 27-June 3, 2018, Gothenburg, Sweden}
\acmPrice{15.00}
\acmDOI{10.1145/3180155.3180197}
\acmISBN{978-1-4503-5638-1/18/05}

\newcommand{\sma}{{\sc SMOTE}}
\newcommand{\smb}{{\sc SMOTUNED}}

\begin{document}

\title{Is ``Better Data'' Better Than ``Better Data Miners''?}
\subtitle{On the Benefits of Tuning SMOTE for Defect Prediction }


\author{Amritanshu Agrawal}
\affiliation{Department of Computer Science\\
North Carolina State University\\
Raleigh, NC, USA\\}
\email{aagrawa8@ncsu.edu}

\author{Tim Menzies}
\affiliation{Department of Computer Science\\
North Carolina State University\\
Raleigh, NC, USA\\}
\email{tim@menzies.us}

\begin{abstract}
We report and fix an important systematic error in prior
studies that ranked classifiers for software analytics.
Those studies  did  not (a)~assess classifiers on multiple   criteria
and they did not 
(b)~study  how variations in the  data affect the results. 
Hence, 
this paper applies (a)  multi-performance criteria while (b)~fixing the weaker regions of the training
 data (using {\smb}, which is an auto-tuning version of {\sma}).
This approach
leads to dramatically large increases in software defect predictions when applied in a 5*5 cross-validation study for  3,681	JAVA classes (containing over a million lines of code) from open source  systems,
{\smb} increased
AUC and recall by 60\% and 20\% respectively. 
These improvements are independent of the classifier used to
predict for defects. Same kind of pattern (improvement) was observed when a comparative analysis of {\sma} and {\smb} was done against the most recent class imbalance technique.

In conclusion, for software analytic tasks like defect prediction, (1)~data
pre-processing can be more important than  classifier
choice,
(2)~ranking studies  are  incomplete  without
 such pre-processing, and
(3)~{\smb} is a   promising candidate for  pre-processing.

\end{abstract}



\maketitle

\section{Introduction}
\label{sect:intro}

Software quality methods cost money and better quality costs exponentially more money ~\cite{voas1995software, fu2016tuning}. Given finite budgets, quality assurance resources are usually 
skewed towards areas known to be most safety critical or mission critical~\cite{lowry1998towards}. This leaves ``blind spots'': regions of the system that may contain defects which may be missed. Therefore, in addition to rigorously assessing  critical areas, a parallel activity should be to {\em sample the blind spots}~\cite{Menzies04}. 

To sample those blind spots, many researchers  use  {\em static code defect predictors}.
Source code is divided into sections and researchers annotate the code with the number of issues known for each section.
Classification algorithms are then applied to learn what static code attributes
distinguish 
between sections with few/many issues.
Such static code measures can be automatically extracted from
the code base with little effort even for very large software
systems \cite{nagappan2005static}.

One perennial problem   is what classifier should be used to build     predictors?
Many papers report {\em ranking studies} where
a quality measure  is collected from    classifiers when they are 
 applied to data sets~\cite{lessmann2008benchmarking,hall2012systematic,elish2008predicting,menzies2010defect,gondra2008applying,radjenovic2013software,jiang2008techniques,wang2013using,mende2009revisiting,li2012sample,khoshgoftaar2010attribute,jiang2009variance,ghotra2015revisiting,jiang2008can,tantithamthavorn2016automated,fu2016tuning,fu2017revisiting}.
These ranking studies report which   classifiers
 generate  best predictors.
 
Research of this paper began with the question {\em would the use of
data pre-processor change the rankings of classifiers?}
SE data
sets are often imbalanced, i.e., the data in the target class is overwhelmed by an over-abundance of information about everything else except the target~\cite{menzies2007problems}.
As shown in the literature review of this paper, in the overwhelming majority of papers (85\%), SE research uses {\sma} to fix data imbalance~\cite{chawla2002smote} but {\sma} is controlled by numerous parameters which
usually are tuned using engineering expertise or left at their default
values. This paper proposes 
{\smb}, an automatic method for setting those parameters which when assessed on  defect data from 3,681	 classes (over a million lines of code) 
taken from open source JAVA systems,  {\smb} out-performed
both the original SMOTE~\cite{chawla2002smote} as well as state-of-the-art method~\cite{bennin2017mahakil}.

To assess, we ask four questions: 
 \bi\item
  \textbf{RQ1}:  {\em Are the default ``off-the-shelf'' parameters for {\sma} appropriate for
  all data sets?} 
  \ei
 \begin{lesson}{\smb} learned different parameters for each data set, all of which  were very different from default {\sma}.
 \end{lesson}
  \bi
  \item
  \textbf{RQ2}: {\em   Is  there any benefit in tuning the default parameters of {\sma} for
  each new data set?} 
  \ei
   \begin{lesson}Performance improvements using {\smb} are dramatically large, e.g., improvements in AUC up to 60\% against {\sma}.
 \end{lesson}
In those results, we see that  while no learner was best across all data sets and   performance criteria,
{\smb} was most often seen in the best results.
That is, creating better training data might be more important
than the subsequent choice of classifiers. 
 
  
   \bi
  \item
  \textbf{RQ3}: {\em  In terms of runtimes, is the cost of running {\smb} worth the performance improvement?}
  \ei
  
   \begin{lesson}{\smb}   terminates in under two minutes, i.e.,  fast enough
   to recommend its widespread use.
 \end{lesson}

   \bi
  \item
  \textbf{RQ4}: {\em  How does {\smb} perform against the recent class imbalance technique?}
  \ei
  
   \begin{lesson}{\smb} performs better than a very recent  imbalance handling technique proposed by Bennin et al.~\cite{bennin2017mahakil}.
 \end{lesson}
 \noindent
 
\noindent
In summary, the  contributions of this paper are:
\bi
\item The discovery of an important systematic error in  many prior ranking studies, i.e., all of
~\cite{lessmann2008benchmarking,hall2012systematic,elish2008predicting,menzies2010defect,gondra2008applying,radjenovic2013software,jiang2008techniques,wang2013using,mende2009revisiting,li2012sample,khoshgoftaar2010attribute,jiang2009variance,ghotra2015revisiting,jiang2008can,tantithamthavorn2016automated,fu2016tuning, fu2017revisiting}.
\item A novel application of search-based SE ({\smb}) to handle class imbalance that out-performs the prior state-of-the-art.
\item Dramatically large improvements in  defect predictors.
\item Potentially, for any other software analytics task that uses classifiers, a way to improve those learners as well.
\item A methodology for assessing the value of pre-processing data sets in software analytics.
\item A reproduction package to reproduce our results then (perhaps) to improve or refute  our results (Available to download from \url{http://tiny.cc/smotuned}). 

\ei
The rest of this paper is structured as follows:
\tion{review} gives an overview on software defect prediction.
\tion{performance} talks about all the performance criteria used in this paper.
\tion{imbalance} explains the problem of class imbalance in defect prediction. Assessment of the previous ranking studies is done in \tion{rank}.
\tion{smote} introduces {\sma} and discusses how {\sma} has been used in literature. \tion{smotuned} provides the definition of {\smb}. \tion{experiment} describes the experimental setup of this paper and above research questions are answered in
\tion{results}. Lastly, we discuss the validity of our results 
and a section describing our conclusions.
 
Note that the experiments of this paper only make conclusions about software analytics for defect prediction. That said,   many other software analytics
tasks use the same classifiers explored here: 
for non-parametric sensitivity analysis~\cite{menzies2000practical},
as a pre-processor to build the tree used to 
infer quality improvement plans~\cite{krishna2017less}, 
to predict Github issue close time~\cite{jones17}, and many more. That is, potentially, {\smb} is  a sub-routine that could improve many software analytics tasks. This could be a highly fruitful direction for future research.




\section{Background and Motivation}

\subsection{Defect Prediction}
\label{sect:review}

Software programmers are intelligent, but busy people. Such busy people often introduce  defects into 
the code they write~\cite{guo2011not}.  Testing software for defects
is expensive and most software assessment budgets are finite. Meanwhile,  assessment effectiveness increases exponentially with assessment effort~\cite{fu2016tuning}. Such exponential costs  exhaust finite resources so software developers
must carefully decide what parts of their code need most testing.

A variety of approaches have been proposed to recognize
 defect-prone  software components using code metrics (lines of code, complexity)~\cite{d2010extensive,menzies2007data, nagappan2006mining,shepperd2014researcher,menzies2010defect} or process metrics (number of changes, recent activity)~\cite{hassan2009predicting}.
Other work, such as that of 
Bird et al.~\cite{bird2009putting}, indicated that it is possible to predict which components (for e.g., modules) are likely locations of
defect occurrence using a component's development history
and dependency structure. 
Prediction models based on the topological properties
of components within them have also  proven to be  
accurate~\cite{zimmermann2008predicting}.

 \begin{table*}[!t]
\renewcommand{\baselinestretch}{0.8}\begin{center}
\caption{OO CK code metrics used for all studies in this paper.
The last line shown, denotes the dependent variable.}
\label{fig:ck}
{\footnotesize
\begin{tabular}{c|l|p{4.4in}}
amc & average method complexity & e.g., number of JAVA byte codes\\
\hline
avg, cc & average McCabe & average McCabe's cyclomatic complexity seen
in class\\
\hline
ca & afferent couplings & how many other classes use the specific
class. \\
\hline
cam & cohesion amongst classes & summation of number of different
types of method parameters in every method divided by a multiplication
of number of different method parameter types in whole class and
number of methods. \\
\hline
cbm &coupling between methods & total number of new/redefined methods
to which all the inherited methods are coupled\\
\hline
cbo & coupling between objects & increased when the methods of one
class access services of another.\\
\hline
ce & efferent couplings & how many other classes is used by the
specific class. \\
\hline
dam & data access & ratio of the number of private (protected)
attributes to the total number of attributes\\
\hline
dit & depth of inheritance tree &\\
\hline
ic & inheritance coupling & number of parent classes to which a given
class is coupled
\\
\hline
lcom & lack of cohesion in methods &number of pairs of methods that do
not share a reference to an case variable.\\
\hline
locm3 & another lack of cohesion measure & if $m,a$ are the number of
$methods,attributes$
in a class number and $\mu(a)$ is the number of methods accessing an
attribute,
then
$lcom3=((\frac{1}{a} \sum, j^a \mu(a, j)) - m)/ (1-m)$.
\\
\hline
loc & lines of code &\\
\hline
max, cc & maximum McCabe & maximum McCabe's cyclomatic complexity seen
in class\\
\hline
mfa & functional abstraction & no. of methods inherited by a class
plus no. of methods accessible by member methods of the
class\\
\hline
moa & aggregation & count of the number of data declarations (class
fields) whose types are user defined classes\\
\hline
noc & number of children &\\
\hline
npm & number of public methods & \\
\hline
rfc & response for a class &number of methods invoked in response to
a message to the object.\\
\hline
wmc & weighted methods per class &\\
\hline
\rowcolor{lightgray}
nDefects & raw defect counts & numeric: number of defects found in post-release bug-tracking systems.\\
\rowcolor{lightgray}
defects present? & boolean& if {\em nDefects} $>0$ then {\em true} else {\em false}
\end{tabular}
}
\end{center}
\vspace{-0.3cm}
\end{table*}

The lesson of all the above  is that the probable location
of future defects can be guessed using   logs of past defects~\cite{hall2012systematic, catal2009systematic}. These logs might
summarize software components using
static code metrics such as 
McCabes  cyclomatic  complexity, Briands coupling metrics, dependencies between  binaries, or
the  CK  metrics~\cite{chidamber1994metrics} (which is described in  Table~\ref{fig:ck}). 
One advantage with CK metrics is that they are  simple  to  compute and hence,
they are widely used. Radjenovi{\'c} et al.~\cite{radjenovic2013software} reported that in
the static code defect prediction, the CK metrics are
used  twice as much (49\%) 
as more traditional source code metrics such as McCabes (27\%) or process metrics (24\%).
The static code measures that can be extracted from a software is shown in   Table~\ref{fig:ck}. Note that such
attributes  can be automatically
collected, even for very large systems~\cite{nagappan2005static}. Other methods,
like manual code reviews, are far slower and far more labor intensive. 

Static code defect predictors are remarkably fast and effective.
Given the current generation of data mining tools, it can be a matter
of just a few seconds to learn a defect predictor (see the runtimes in Table~9 of reference~\cite{fu2016tuning}). Further, in a recent study by Rahman et
al.~\cite{Rahman14}, found no significant differences in the cost-effectiveness
of
(a) static code analysis tools FindBugs and Jlint, and (b) static code defect predictors.
This is an interesting result since it is  much slower to adapt static code
analyzers to new  languages than defect predictors (since the latter just requires hacking together some new
static code metrics extractors).

\subsection{Performance Criteria}
\label{sect:performance}

Formally, defect prediction is a binary classification problem.
The performance of a defect predictor can be assessed via a  confusion matrix like Table~\ref{fig:cmatrix}
where a ``positive'' output is the defective class  under study and a ``negative'' output is the non-defective one.

\begin{wraptable}{r}{1.5in}
\small
\begin{center}
\vspace{-0.5cm}
\caption{Results Matrix} 
\vspace{-0.3cm}
\label{fig:cmatrix}
\begin{tabular} {@{}cc|c|c|l@{}}
\cline{3-4}
& & \multicolumn{2}{ c| }{Actual} \\ \cline{2-4}
& \multicolumn{1}{ |c| }{Prediction} & false & true  \\ \cline{2-4}
& \multicolumn{1}{ |c| }{defect-free} & $\mathit{TN}$ & $\mathit{FN}$ & \\ \cline{2-4}
 & \multicolumn{1}{ |c| }{defective} & $\mathit{FP}$& $\mathit{TP}$  &  \\ \cline{2-4}
\cline{2-4}
\end{tabular}
\end{center} 
\vspace{-0.3cm}
\end{wraptable}

Further, ``false'' means the learner got it wrong and ``true'' means the learner correctly identified
a fault or non-fault module. Hence, Table~\ref{fig:cmatrix} has four quadrants containing, e.g., $\mathit{FP}$ which denotes ``false positive''.

From this matrix, we can define performance measures like: 
\bi
\item \textbf{Recall} $=$ $pd$  $=$ $\mathit{TP}/(\mathit{TP} + \mathit{FN})$

\item  \textbf{Precision}  $=$ $prec$ $=$ $\mathit{TP}/(\mathit{TP} + \mathit{FP})$

\item \textbf{False Alarm}  $=$ $pf$ $=$ $\mathit{FP}/(\mathit{FP} + \mathit{TN})$

\item \textbf{Area Under Curve (AUC)}, which 
is the area covered by an ROC curve~\cite{swets1988measuring, duda2012pattern} in which the X-axis represents, false positive rate and the Y-axis
represents true positive rate.
\ei
 
\begin{wrapfigure}{r}{2.0in}
 \vspace{-0.4cm}
\begin{center}
\includegraphics[width=2.0in]{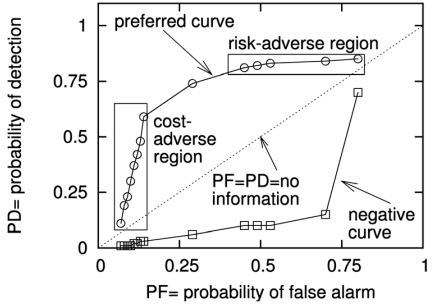}
\end{center}
\vspace{-0.5cm}
\caption{Trade-offs false alarm vs
recall (probability of detection).  }\label{fig:trade}
\vspace{-0.3cm}
 \end{wrapfigure}

As shown in Figure~\ref{fig:trade},
a typical predictor must ``trade-off''
between false alarm and recall.
This is because the  more sensitive the detector, the more often it triggers and the higher its recall. If a detector triggers more often, it also raises more false alarms.
Hence, when increasing recall, we  should  expect
the false alarm rate to  increase
(ideally, not by very much).
 
There are many more ways to evaluate defect predictors besides the four listed above.
Previously, Menzies et al. catalogued dozens of them (see Table 23.2 of~\cite{menzies2014sharing}) and even several novel ones were proposed
(balance, G-measure~\cite{menzies2007data}).  
But no evaluation criteria is ``best'' since different  criteria are appropriate in different business contexts. For e.g., as shown
in 
Figure~\ref{fig:trade},
when dealing
with safety-critical applications, management may be
``risk adverse'' and hence many elect
 to {\em maximize recall}, regardless of the time wasted exploring  false alarm.
 Similarly, 
when rushing some non-safety critical application to market, management may be ``cost adverse''
and elect to {\em minimize false alarm} since this avoids distractions to the developers. 

In summary, there are  numerous evaluation criteria and  numerous business contexts
where different criteria might be preferred by different local business users. In response to
the cornucopia of evaluation criteria, we make the following recommendations: a) do evaluate learners on more than one criteria, b) do not evaluate learners on  all criteria (there are too many), and instead, apply the criteria widely seen in the literature. Applying this advice, this paper  evaluates the defect predictors using the four criteria
mentioned above (since these are widely reported in the literature~\cite{ghotra2015revisiting,fu2016tuning})) but not other 
 criteria  that  have yet to gain a wide acceptance 
 (i.e., balance and G-measure).
 

\subsection{Defect Prediction and Class Imbalance}
\label{sect:imbalance}
  Class imbalance  is concerned with the situation in where some classes of data are
highly under-represented compared to other classes~\cite{he2009learning}.
By convention,
the under-represented class is called the {\em minority} class,
and correspondingly the class which is over-represented is called the
{\em majority} class. In this paper, we say that class imbalance is {\em worse}
when the ratio of minority class to majority {\em increases}, that is,
{\em class-imbalance of 5:95} is worse than {\em 20:80}. Menzies et al.~\cite{menzies2007problems} reported SE data sets often contain class imbalance. In their examples, they showed static code defect prediction data sets with
class imbalances of 1:7; 1:9; 1:10; 1:13; 1:16; 1:249.

 \begin{table*}[!t]
 \caption{Classifiers used in this study.
 Rankings
 from Ghotra et al.~\cite{ghotra2015revisiting}.}
 \vspace{-0.2cm}
 \label{tbl:learners}
 \footnotesize
 \begin{tabular}{l|l|p{4.5in}}
{\bf RANK} & {\bf LEARNER} & {\bf NOTES}\\\hline
 1 ``best'' & RF= random forest & 
 Random forest of entropy-based decision trees.\\\cline{2-3}
 &  LR=Logistic regression &
 A generalized linear regression
model.\\\hline
 2 & KNN= K-means &  Classify a new instance by finding ``k'' examples of similar instances.
 Ghortra et al. suggested
 $K=8$.\\\cline{2-3}
 & NB= Naive Bayes &  Classify a new instance by (a)~collecting mean and standard deviations of attributes in old instances of  different classes; (b)~return the class whose attributes are statistically most similar to the new instance.\\\hline
 3 & DT= decision trees & Recursively
 divide data by selecting attribute splits
 that reduce the entropy of the class distribution.\\\hline

 4 ``worst'' & SVM= support vector machines &
 Map the raw data into a higher-dimensional space where it is easier to distinguish the examples.
 \\\hline
 \end{tabular}
 \vspace{-0.2cm}
 \end{table*}
 
  \begin{table}[!t]
\centering
\footnotesize
\caption{22 highly cited Software Defect prediction studies.}
\vspace{-0.2cm}
\label{tbl:survey2}
    \begin{tabular}{c|c|c|c|c|c}
        \begin{tabular}[c]{@{}c@{}}\textbf{Ref}\end{tabular} & \textbf{Year} & \textbf{Citations} & \begin{tabular}[c]{@{}c@{}}\textbf{Ranked} \\\textbf{Classifiers?} \end{tabular} &\begin{tabular}[c]{@{}c@{}} \textbf{Evaluated} \\\textbf{using} \\\textbf{multiple} \\\textbf{criteria?}\end{tabular}&\begin{tabular}[c]{@{}c@{}} \textbf{Considered}\\\textbf{Data}\\\textbf{Imbalance?} \end{tabular}\\ \hline
        \cite{menzies2007data} & 2007 & 855 & \cmark & 2 & \xmark \\
        \cite{lessmann2008benchmarking} & 2008 & 607 & \cmark & 1 & \xmark \\ 
        \cite{elish2008predicting} & 2008 & 298 & \cmark & 2 & \xmark\\  
        \cite{menzies2010defect} & 2010 & 178 & \cmark & 3 & \xmark \\  
        \cite{gondra2008applying} & 2008 & 159 & \cmark & 1 & \xmark\\     
        \cite{kim2011dealing} & 2011 & 153 &\cmark & 2 & \xmark \\ 
        \cite{radjenovic2013software} & 2013 & 150 & \cmark & 1 & \xmark \\   
        \cite{jiang2008techniques} & 2008 & 133 & \cmark & 1 & \xmark \\    
        \cite{wang2013using} & 2013 & 115 & \cmark & 1 & \cmark \\  
        \cite{mende2009revisiting} & 2009 & 92 & \cmark & 1 & \xmark \\          
        \cite{li2012sample} & 2012 & 79 & \cmark & 2 & \xmark  \\ 
        \cite{kamei2007effects} & 2007 & 73 & \xmark & 2 & \cmark\\  
        \cite{pelayo2007applying} & 2007 & 66 & \xmark & 1 & \cmark \\  
        \cite{jiang2009variance} & 2009 & 62 & \cmark & 3 & \xmark  \\ 
        \cite{khoshgoftaar2010attribute} & 2010 & 60 & \cmark & 1 & \cmark  \\  
        \cite{ghotra2015revisiting} & 2015 & 53 & \cmark & 1 & \xmark  \\  
        \cite{jiang2008can} & 2008 & 41 & \cmark & 1 & \xmark  \\  
         \cite{tantithamthavorn2016automated} & 2016 & 31 & \cmark & 1 & \xmark  \\ 
        \cite{tan2015online} & 2015 & 27 & \xmark & 2 & \cmark \\  
        \cite{pelayo2012evaluating} & 2012 & 23 & \xmark & 1 & \cmark \\  
        \cite{fu2016tuning} & 2016 & 15 & \cmark & 1 & \xmark  \\  
        \cite{bennin2017mahakil} & 2017 & 0 & \cmark & 3 & \cmark \\
\end{tabular}
\vspace{-0.3cm}
\end{table}
 
The problem of class imbalance is sometimes discussed in the software analytics community.
Hall et al.~\cite{hall2012systematic} found that models based on C4.5 under-perform if they have imbalanced data while Naive Bayes and Logistic regression perform relatively better. 
Their general recommendation is to not use
imbalanced data.  
Some researchers offer preliminary explorations into methods that might mitigate for class imbalance.
Wang et al.~\cite{wang2013using} and Yu et al.~\cite{yu2017performance} validated the Hall et al. results and concluded that the
performance of C4.5 is unstable on imbalanced data sets while  Random Forest and Naive Bayes are 
more stable. 
Yan et al.~\cite{yan2010software} performed fuzzy logic and rules to overcome the imbalance problem, but they only
explored one kind of learner (Support Vector Machines).
Pelayo et al.~\cite{pelayo2007applying} studied the effects of the percentage of oversampling and undersampling done. They found out that different percentage of each helps improve the accuracies of decision tree learner for defect prediction using CK metrics. Menzies et al.~\cite{menzies2008implications} undersampled the non-defect class to balance training
data and reported how little information was required to learn a defect predictor. They found that throwing away data does not degrade the performance of Naive Bayes and C4.5 decision trees. Other papers~\cite{pelayo2007applying, pelayo2012evaluating, riquelme2008finding} have shown the usefulness of resampling based on different learners.

We note that  
many researchers in this area~\cite{gray2009using,yu2017performance,wang2013using} refer to the {\sma} method explored in this paper,  but only in the context of future work.
 One rare exception to this general pattern is the recent paper by   Bennin et al.~\cite{bennin2017mahakil}, which we explored as part of
RQ4.


 \subsection{Ranking Studies}
\label{sect:rank}

A constant problem in defect prediction is what  classifier should be applied to  build  the  defect  predictors?
To address this problem, many researchers run {\em ranking studies} where  performance scores 
are collected from  many classifiers  executed on  many software defect data sets~\cite{lessmann2008benchmarking,hall2012systematic,elish2008predicting,menzies2010defect,gondra2008applying,radjenovic2013software,jiang2008techniques,wang2013using,mende2009revisiting,li2012sample,khoshgoftaar2010attribute,jiang2009variance,ghotra2015revisiting,jiang2008can,tantithamthavorn2016automated,fu2016tuning}.
This section assesses those ranking studies. We will say a ranking study is ``good'' if it compares multiple learners using multiple data sets and multiple evaluation criteria
while at the same time doing something to address the data imbalance problem.

 \begin{wrapfigure}{r}{1.8in}
  \centering
  \captionsetup{justification=centering}
  \includegraphics[width=1.8in]{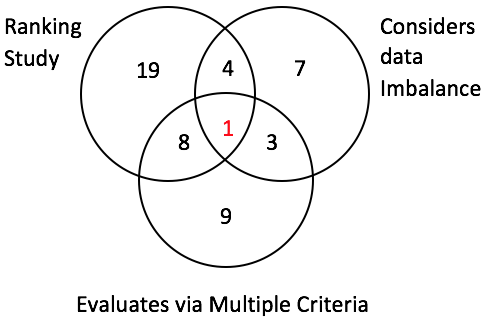}
  \caption{Summary of  Table~\ref{tbl:survey2}.}
\label{fig:s2}
\vspace{-0.4cm}
\end{wrapfigure}
In July 2017,  we searched
scholar.google.com for the conjunction of ``software'' and ``defect prediction'' and ``OO'' and ``CK'' published in the last decade. This returned 231 results.
We only selected OO and CK keywords since CK metrics are more popular and better than process metrics for software defect prediction~\cite{radjenovic2013software}.
From that list, we selected ``highly-cited'' papers, which we defined as having more than 10 citations per year. This reduced our population of papers down to 107.
After reading the titles and abstracts of those papers, and skimming the contents of the potentially interesting papers, we found 22 papers of Table~\ref{tbl:survey2} that either performed ranking studies
(as defined above) or studied the effects of class imbalance on defect prediction. In the column ``evaluated using
multiple criteria'',
papers scored more than ``1'' if they used multiple performance scores  of the kind listed at the end of \tion{performance}.

We find that, in those 22 papers from Table~\ref{tbl:survey2},
numerous classifiers have used AUC as the measure to evaluate the software defect
predictor studies. We also found that majority of papers (from last column of Table~\ref{tbl:survey2}, 6/7=85\%) in SE community has used SMOTE to fix the data imbalance~\cite{wang2013using,kamei2007effects,pelayo2007applying,tan2015online,pelayo2012evaluating,bennin2017mahakil}. This also made us to propose {\smb}. As
 noted in~\cite{lessmann2008benchmarking, ghotra2015revisiting},  no single classification technique always dominates.  
That said, Table~IX of a recent study by Ghotra et al.~\cite{ghotra2015revisiting} ranks
 numerous classifiers  using data similar
 to what we use here (i.e., OO JAVA systems described using CK metrics).
  Using their work, we can
select
 a range of classifiers  for this study
 ranking from ``best''
 to ``worst': see Table~\ref{tbl:learners}.

 



The key observation to be made from  this 
survey is that, as shown in Figure~\ref{fig:s2}, the overwhelming majority of
prior papers in our sample {\em do not satisfy}  our definition of a ``good'' project
(the sole exception is the  recent   Bennin et al.~\cite{bennin2017mahakil} which we explore in RQ4).
Accordingly, the rest of this
paper defines and executes a ``good'' ranking  study, with an additional
unique feature of   an auto-tuning version of {\sma}.

\subsection{Handling Data Imbalance with SMOTE}
\label{sect:smote}

{\sma} handles class imbalance by changing the frequency of different classes of the training
data~\cite{chawla2002smote}. 
The algorithm's name is short for ``synthetic minority over-sampling technique''.
When applied to data, {\sma} sub-samples the majority class (i.e., deletes some examples)
while super-sampling the minority class
until
all classes have the same frequency.  In the case of software defect data,
the minority class is usually the  defective class.

 \begin{wrapfigure}{r}{2in}\small
\begin{lstlisting}[mathescape,linewidth=2in,frame=n,numbers=none]
  def SMOTE(k=2, m=50%, r=2):  # defaults
    while Majority > m do
      delete any majority item # random
    while Minority < m do
      add something_like(any minority item)
      
  def something_like(X0): 
    relevant = emptySet
    k1 = 0
    while(k1++ < 20 and size(found) < k)  {
       all = k1 nearest neighbors
       relevant += items in "all" of X0 class}
    Z = any of found
    Y = interpolate (X0, Z)
    return Y
    
  def minkowski_distance(a,b,r): 
    return $( \Sigma_i\ abs(a_i - b_i)^r)^{1/r}$
\end{lstlisting} 
\vspace{-0.2cm}
\caption{Pseudocode of SMOTE}
\label{fig:pseudocode} 
\vspace{-0.4cm}
\end{wrapfigure}

Figure~\ref{fig:pseudocode} shows how {\sma} works. During super-sampling,
a member of the minority class finds $k$ nearest neighbors. It builds an artificial member
of the minority class at some point in-between itself and one of its random nearest
neighbors.  During that process, some distance function is required which is the {\em minkowski\_distance} function. 

{\sma}'s control parameters are (a) $k$ that selects how many neighbors to use  (defaults to $k=5$), (b) $m$ is how many examples of each class which need to be generated (defaults to $m=50\%$ of the total training samples), and (3) $r$ which selects  the distance function (default is $r=2$,
i.e., use    
  Euclidean distance).


In the software analytics literature, there are contradictory findings on
the value of applying {\sma} for software defect prediction.
Van et al.~\cite{van2007experimental}, Pears et al.~\cite{pears2014synthetic} and Tan et al.~\cite{tan2015online} found {\sma} to be advantageous, while others, such as Pelayo et al.~\cite{pelayo2007applying} did not.

Further, some researchers report that some learners respond better than others to {\sma}. Kamei et al.~\cite{kamei2007effects} evaluated the effects of {\sma} applied to  four fault-proneness models
(linear discriminant analysis, logistic regression, neural network, and decision tree) by
using two module sets of industry legacy software. They reported that {\sma} improved the prediction performance of the linear and logistic models, but not neural network and decision tree models. Similar results, that the value of {\sma} was dependent on the learner,
was also reported by Van et al.~\cite{van2007experimental}.

Recently, Bennin et al.~\cite{bennin2017mahakil}  proposed a new method based on the chromosomal theory of inheritance. 
Their MAHAKIL algorithm interprets two distinct sub-classes as parents and generates a new synthetic instance that inherits different traits from each parent and contributes to the diversity within the data distribution.
They report that MAHAKIL usually performs as well as  {\sma}, but
does much better than all   other class balancing techniques in terms of recall.
Please note, that work did not  consider the impact of parameter tuning of a preprocessor so in our RQ4 we will compare {\smb} to MAHAKIL.

\vspace{-0.2cm}
\subsection{SMOTUNED = auto-tuning {\sma}}
\label{sect:smotuned}

One possible explanation for the variability in the {\sma} results is that the
default parameters of this algorithm are not suited to all data sets. To test this,
we designed {\smb}, which is an auto-tuning version of {\sma}. {\smb}
uses different control parameters for different data sets.

{\smb} uses DE (differential evolution~\cite{storn1997differential}) to explore the parameter space of
Table~\ref{tb:tuned}.  DE is an
optimizer useful for functions that may not be smooth or linear.  Vesterstrom et al.~\cite{vesterstrom2004comparative} find   DE's optimizations to be  competitive with other optimizers like 
   particle swarm optimization or genetic algorithms.
   DEs have been used before for   parameter tuning~\cite{omran2005differential, chiha2012tuning,fu2016tuning,fu2017easy, agrawal2016wrong}) but this paper is the first attempt to do
   DE-based class re-balancing for SE data by studying multiple learners for multiple evaluation criteria.

In Figure~\ref{fig:pseudo_DE}, DE evolves a \textit{frontier} of
candidates from an initial population which is driven by a goal (like maximizing recall) evaluated using a fitness function (shown in line 17). In the case of {\smb},
each  candidate is a randomly selected value for {\sma}'s $k, m$ and $r$ parameters.
 To evolve the frontier, within each generation,
 DE compares each item to a {\em new} candidate generated
 by combining three other frontier items (and better {\em new} candidates replace
 older items). 
 To compare them, the {\em better} function (line 17) calls $SMOTE$ function (from Figure~\ref{fig:pseudocode}) using the proposed {\em new} parameter settings. This pre-processed training data is then fed into a classifier to find a particular measure (like recall).
 When our DE  terminates, it returns the best candidate ever seen in the entire run.
 
Table~\ref{tb:algo2} provides important terms of {\smb} when exploring SMOTE's
  parameter ranges,  shown  in  Table~\ref{tb:tuned}. To define
the parameters, we found the range of used settings for {\sma} and distance functions
in the   SE and machine learning  literature. To avoid introducing noise by overpopulating the minority samples we are not using $m$ as percentage rather than number of examples to create. Aggarawal et al.~\cite{aggarwal2001surprising}
argue that with data being highly dimensional, $r$ should shrink to some fraction less than one
(hence the bound of $r=0.1$ in Table~\ref{tb:tuned}). 

\begin{figure}[!t]
\small 
\begin{lstlisting}[mathescape,linewidth=6.7cm,frame=none,numbers=right ]
  def DE( n=10, cf=0.3, f=0.7):  # default settings
    frontier = sets of guesses (n=10)
    best = frontier.1 # any value at all
    lives = 1
    while(lives$--$ > 0): 
      tmp = empty
      for i = 1 to $|$frontier$|$: # size of frontier
         old = frontier$_i$
         x,y,z = any three from frontier, picked at random
         new= copy(old)  
         for j = 1 to $|$new$|$: # for all attributes
           if rand() < cf    # at probability cf...
              new.j = $x.j + f*(z.j - y.j)$  # ...change item j
         # end for
         new  = new if better(new,old) else old
         tmp$_i$ = new 
         if better(new,best) then
            best = new
            lives++ # enable one more generation
         end                  
      # end for
     frontier = tmp
    # end while
    return best
\end{lstlisting} 
\caption{SMOTUNED uses DE (differential evolution).}
\label{fig:pseudo_DE} 
\vspace{-0.3cm}
\end{figure}

\begin{table}[!t]
    \begin{center}
    \footnotesize
\caption{SMOTE parameters}
\label{tb:tuned}
 \begin{tabular}{p{0.5cm}|c@{~}|c@{~}|p{3.3cm}}
         \textbf{Para} &  \begin{tabular}[c]{@{}c@{}}\textbf{Defaults} \\ \textbf{used by} \\\textbf{{\sma}} \end{tabular}
          &  \begin{tabular}[c]{@{}c@{}}
          \textbf{Tuning Range} \\
           \textbf{(Explored by} \\ \textbf{( {\smb})} \end{tabular}&  \textbf{Description} \\
        \hline
        $k$ & 5 & [1,20] & Number of neighbors \\ 
        \hline
       $m$ & 50\% & \{50, 100, 200, 400\} & Number of synthetic examples to create. Expressed as a percent  of final training data. \\ 
        \hline
        $r$ & 2 & [0.1,5] & Power parameter for the Minkowski distance metric.\\
 
\end{tabular}
\end{center}
\vspace{-0.3cm}
\end{table}

\begin{table}[!t]
\begin{center}
\footnotesize
\caption{Important Terms of SMOTUNED Algorithm}
\label{tb:algo2}
\begin{tabular}{|c@{~}|p{4.0cm}|}
        \hline 
        \textbf{Keywords} & \textbf{Description}\\
        \hline

        \begin{tabular}[c]{@{}c@{}}Differential weight $(f=0.7)$\end{tabular} & Mutation power\\
        \hline
        \begin{tabular}[c]{@{}c@{}}Crossover probability $(cf=0.3)$\end{tabular} & Survival of the candidate\\
        \hline
        \begin{tabular}[c]{@{}c@{}}Population Size $(n=10)$\end{tabular} &  Frontier size in a generation \\
        \hline
        \begin{tabular}[c]{@{}c@{}}Lives\end{tabular} & Number of generations\\
        \hline
        \begin{tabular}[c]{@{}c@{}}Fitness Function $(better)$\end{tabular} & Driving factor of DE\\
        \hline
        \begin{tabular}[c]{@{}c@{}}Rand() function\end{tabular} & Returns between 0 to 1 \\
        \hline
        \begin{tabular}[c]{@{}c@{}}Best (or Output)\end{tabular} & Optimal configuration for {\sma} \\
        \hline
\end{tabular}
\end{center}
\vspace{-0.3cm}
\end{table}

\section{Experimental Design}
\label{sect:experiment}
 
This experiment  reports the effects on   defect prediction
after using MAHAKIL or {\smb} or {\sma}. Using some data $D_i \in D$, performance measure $M_i \in M$, and classifier $C_i \in C$,
this experiment conducts the 5*5 cross-validation study, defined below.
Our data sets  $D$ are shown in  Table~\ref{tb:dataset}. These are all open source
JAVA OO systems described in terms of the CK metrics. 
Since, we are comparing these results for imbalanced class, only imbalanced class data sets were selected from SEACRAFT (\url{http://tiny.cc/seacraft}).

Our performance measures $M$ were introduced in \tion{performance}
which includes   AUC, precision, recall, and the  false alarm. 
Our classifiers
 $C$  come from a  recent study~\cite{ghotra2015revisiting}
and were listed in  Table~\ref{tbl:learners}.
For  implementations 
of these learners,
we used  the open source tool
Scikit-Learn~\cite{pedregosa2011scikit}.
Our  cross-validation study~\cite{refaeilzadeh2009cross} is defined as follows:
\be
\item We randomized the order of the data set $D_i$  five times. This reduces the probability
that some random ordering of examples in the data will conflate our results.
\item Each time, we divided the data $D_i$ into five bins;
\item For each bin (the test), we trained on four bins (the rest) and then tested
on the test bin as follows.
\be
\item
The  training set is pre-filtered using either No-SMOTE (i.e., do nothing) or  {\sma} or {\smb}.  
\item
When using {\smb}, we further divide those four bins of training data. 3 bins are used for training the model, and 1 bin is used for validation in DE. DE is  run to  improve
the performance measure $M_i$ seen when the classifier $C_i$ was applied to the training data.
Important point: {\em we only used {\sma} on the training data,  leaving
the  testing data unchanged}.
\item
After pre-filtering, a classifier $C_i$  learns a predictor.
\item
The model is applied to the test data to collect performance measure $M_i$. 
\item 
We print the {\em relative performance delta} between this $M_i$ and another  $M_i$
generated from applying $C_i$ to the raw data $D_i$ (i.e., compare the learner
without any filtering). We finally report median on the 25 repeats.
\ee
\ee

 \begin{table}[!b]
\begin{center}
\small
\caption{Data set statistics. Data sets are sorted from low percentage of defective class to high defective class.
Data comes from the SEACRAFT repository: http://tiny.cc/seacraft}.
\vspace{-0.2cm}
\label{tb:dataset}

\begin{tabular}{r@{~}|r@{~}|r@{~}|r@{~}|r@{~}}
\footnotesize
        \begin{tabular}[c]{@{}c@{}}\textbf{Version}\end{tabular} & \begin{tabular}[c]{@{}c@{}} \textbf{Dataset Name}\end{tabular} & \textbf{Defect \%} & \begin{tabular}[c]{@{}c@{}} \textbf{No. of classes}\end{tabular}&\begin{tabular}[c]{@{}c@{}} \textbf{lines of code} \end{tabular}\\ \hline
4.3 & jEdit & 2  & 492 & 202,363 \\ 
1.0 &   Camel & 4  & 339 & 33,721 \\  
6.0.3 &   Tomcat & 9  & 858 & 300,674\\ 
2.0 &   Ivy & 11 & 352 & 87,769 \\  
1.0 & Arcilook & 11.5  & 234 & 31,342\\ 
1.0 & Redaktor & 15 & 176 & 59,280 \\ 
1.7 & Apache Ant & 22 & 745 & 208,653 \\  
1.2 &   Synapse & 33.5 & 256 & 53,500 \\ 
1.6.1 &   Velocity & 34 & 229 & 57,012 \\\hline
  \multicolumn{3}{r|}{ total:} &    3,681&	1,034,314\\ 

\end{tabular}
\end{center} 
\end{table}

\begin{figure*}[!t]
    \centering
    \begin{minipage}{.33\textwidth}
    \centering
    \includegraphics[width=.95\linewidth]{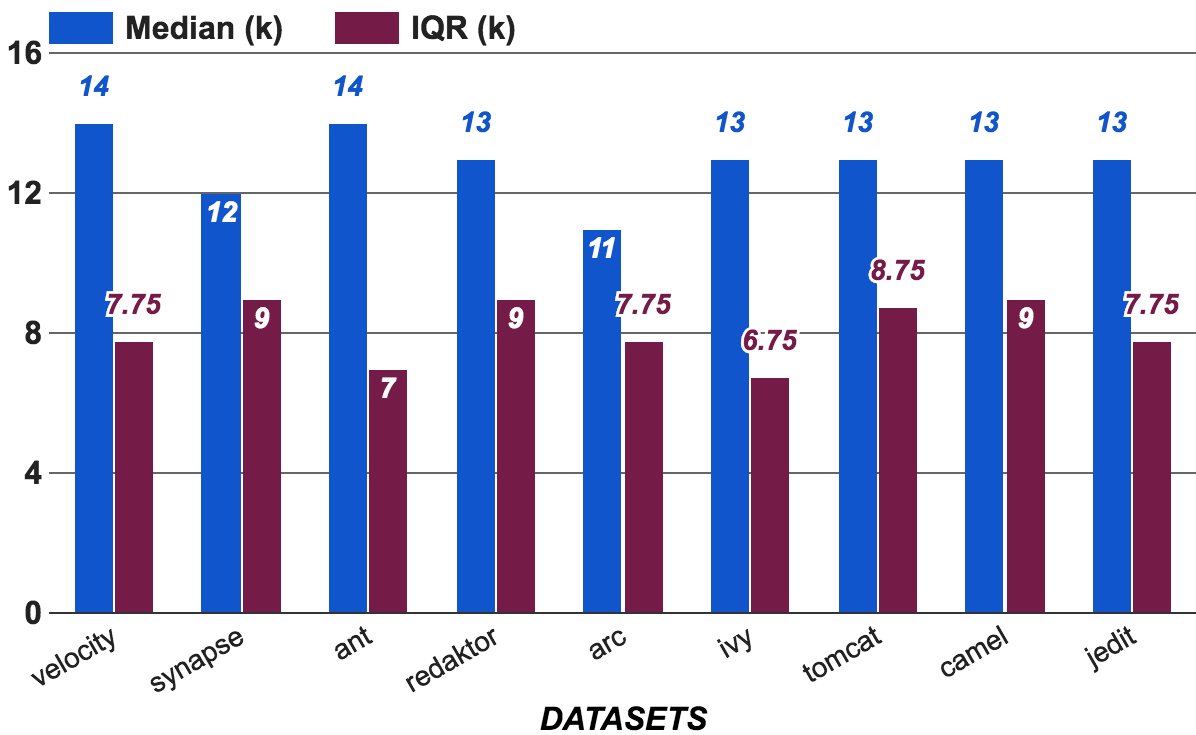}
        {\bf Figure~\ref{fig:para}a:} Tuned values for $k$\\ (default:  $k=5$).
    \end{minipage}~~%
    \begin{minipage}{.33\textwidth}
    \centering
        \includegraphics[width=.95\linewidth]{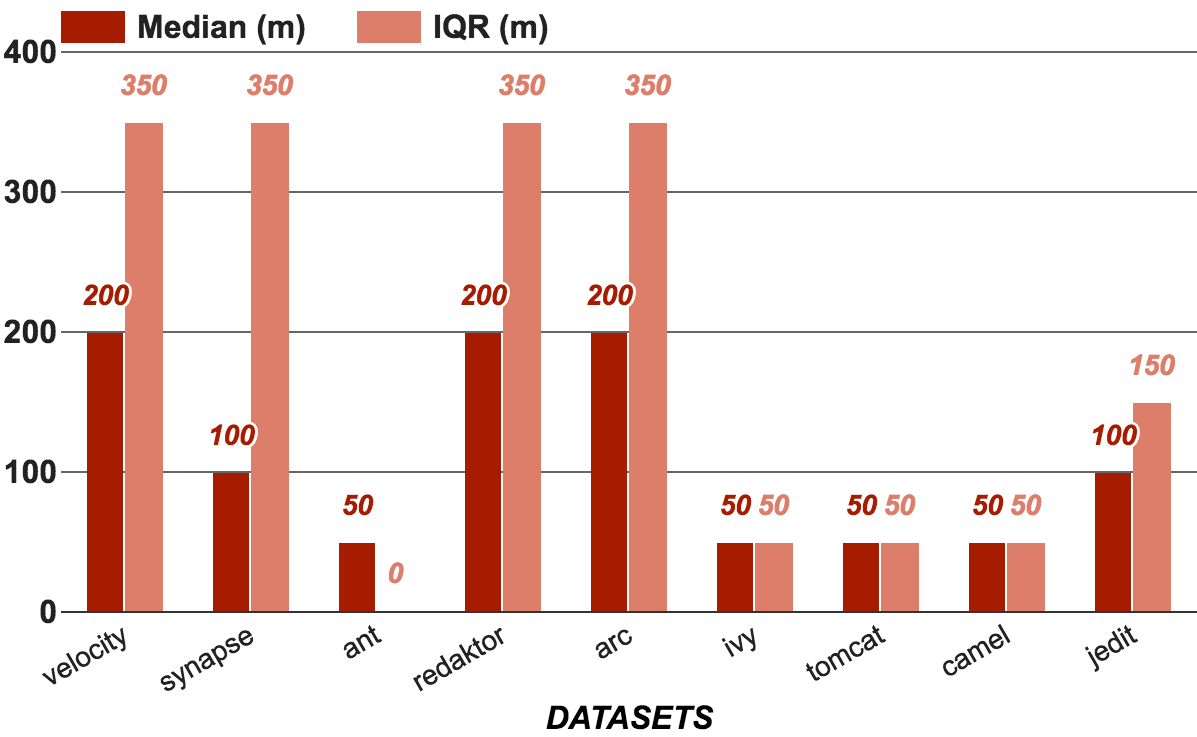}
        {\bf Figure~\ref{fig:para}b:} Tuned values for $m$\\ (default: $m=50\%$).
    \end{minipage}~~%
    \begin{minipage}{.33\textwidth}
    \centering
        \includegraphics[width=.95\linewidth]{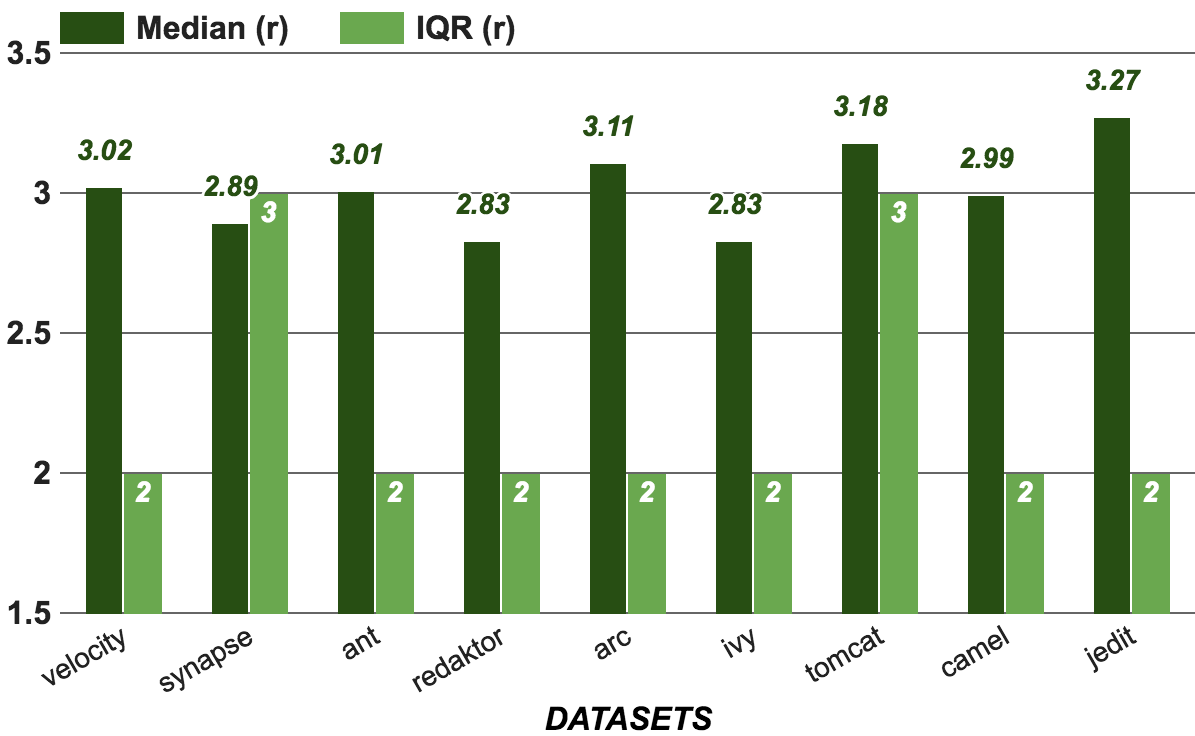}
        {\bf Figure~\ref{fig:para}c:} Tuned values for $r$\\ (default:  $r=2$).
    \end{minipage}
    \vspace{-0.2cm}
    \caption{Data sets vs Parameter Variation when optimized for recall and results reported on recall.
    ``Median'' denotes 50th percentile values seen in the 5*5 cross-validations and ``IQR'' shows the intra-quartile
    range, i.e., (75-25)th percentiles.}
    \label{fig:para}
\end{figure*}

\begin{figure*}[!t]
\begin{minipage}{.5\linewidth}
\centering
 
        \includegraphics[width=0.75\linewidth,keepaspectratio,trim=0cm 1cm 0cm 2cm]{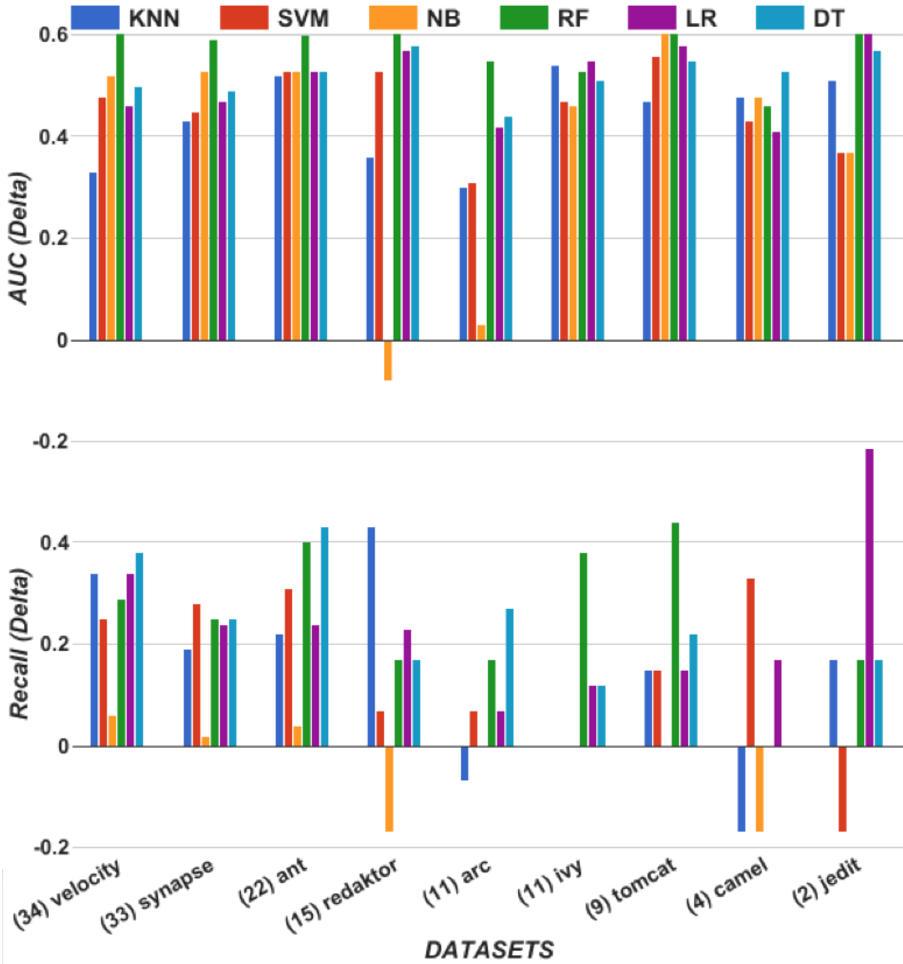}
    \end{minipage}%
\begin{minipage}{.5\linewidth}
        \centering
        \includegraphics[width=0.75\linewidth,keepaspectratio,trim=0cm 1cm 0cm 2cm]{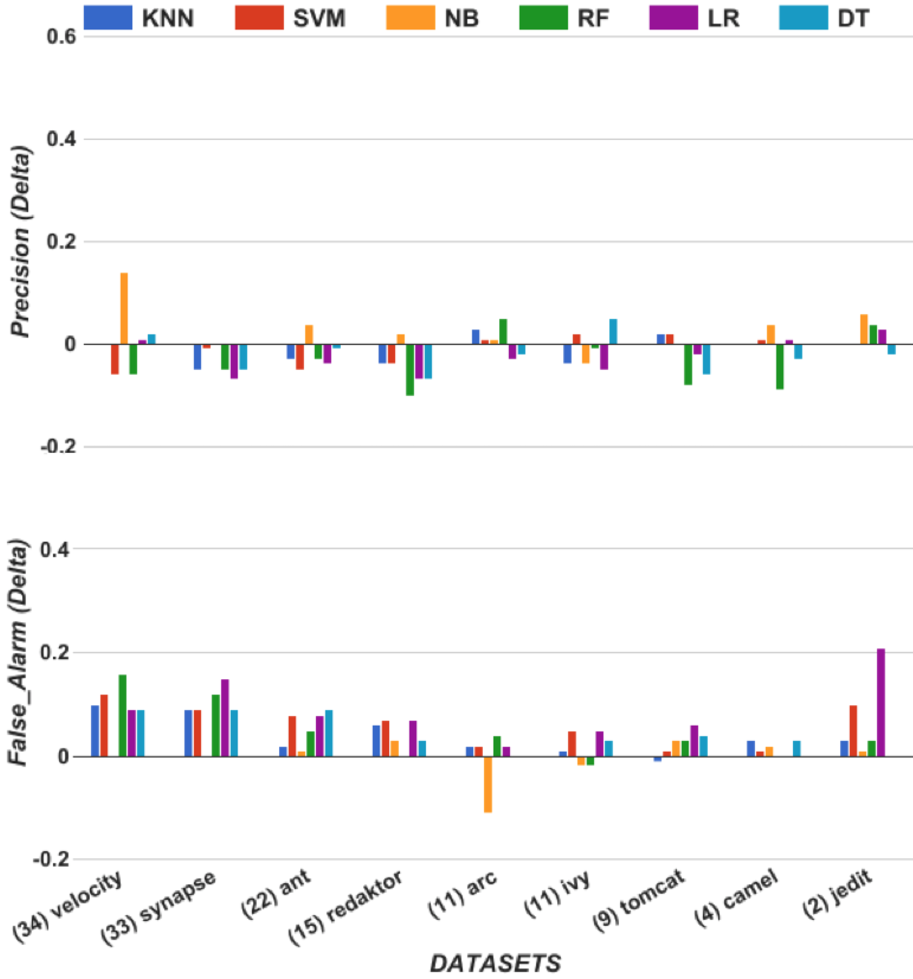}
    \end{minipage}%
    
    \caption{{\smb} improvements over {\sma}. \underline{Within}-Measure
    assessment (i.e., for each of these charts,
    optimize for performance measure $M_i$, then test for
    performance measure $M_i$). For most charts,
    {\em larger} values are {\em better}, but for false alarm,
    {\em smaller} values are {\em better}. Note that the corresponding percentage of minority class (in this case, defective class) is written beside each data set.}
    \label{fig:tuned}
    \vspace{-0.3cm}
\end{figure*}

Note that the above rig tunes {\sma}, but not the control parameters of the classifiers.
We do this since, in this paper,  we aim to document the   benefits of tuning {\sma} since as shown below, they are very large indeed. Also, it would be very useful if we can show that a single algorithm ({\smb})  improves the performance of defect prediction. This would allow
subsequent work to focus on the task of optimizing  {\smb} (which would be a far easier
task than optimizing the tuning of a wide-range of classifiers).

\subsection{Within- vs Cross-Measure Assessment}
\label{sect:wcm}

We call the above rig as the {\em within-measure assessment rig} since it is  biased in its evaluation measures. Specifically,  in this rig,
when {\smb} is optimized for (e.g.,) AUC, we do not explore the effects on (e.g.,) the false alarm. This is less than ideal
since it is known that our performance measures are inter-connected via the Zhang's equation~\cite{zhang2007comments}. Hence, increasing (e.g.,) recall might potentially have the adverse
effect of  driving up (e.g) the false alarm rate. 
To avoid this problem, we also apply the following {\em cross-measure assessment rig}.
At the conclusion of the {\em within-measure assessment rig}, we will observe  that the AUC performance measure will show the largest improvements. Using that best performer, we will re-apply steps 1,2,3 abcde (listed above) but this time:
\bi
\item In step 3b, we will tell {\smb} to optimize for AUC;
\item In step 3d, 3e we will collect the performance delta on AUC as well as precision, recall,
and false alarm.
\ei
In this approach, steps 3d and 3e collect the information required   to check if succeeding according to one performance criteria results in damage to another. We would also want to make sure that our model is not over-fitted based on one evaluation measure. And since {\smb} is a time expensive task, we do not want to tune for each measure which will quadruple the time. The results of within- vs cross-measure assessment is shown in Section \ref{sect:results}.

\subsection{Statistical Analysis}
\label{sec:scott-knott}

When comparing the results of {\smb} to other
treatments, we use a statistical
significance test and an effect size test.
Significance test are useful for detecting if two populations
differ merely by random noise. 
Also, effect sizes are useful for checking that two populations differ by more than just a trivial amount.

For the significance test,  we used the 
     Scott-Knott procedure~\cite{mittas2013ranking,ghotra2015revisiting}. This
     technique recursively bi-clusters a sorted
    set of numbers. If any two clusters are statistically indistinguishable, Scott-Knott
    reports them both as one group.
    Scott-Knott first looks for a break in the sequence that maximizes the expected
    values in the difference in the means before and after the break.
    More specifically,  it  splits $l$ values into sub-lists $m$ and $n$ in order to maximize the expected value of differences  in the observed performances before and after divisions. For e.g., lists $l,m$ and $n$ of size $ls,ms$ and $ns$ where $l=m\cup n$, Scott-Knott divides the sequence at the break that maximizes:
     \[E(\Delta)=ms/ls*abs(m.\mu - l.\mu)^2 + ns/ls*abs(n.\mu - l.\mu)^2\]
Scott-Knott then applies some statistical hypothesis test $H$ to check if $m$ and $n$ are significantly different. If so, Scott-Knott then recurses on each division.
    For this study, our hypothesis test $H$ was a conjunction of the A12 effect size test (endorsed by
    \cite{arcuri2011practical})  and non-parametric bootstrap sampling \cite{efron94}, i.e., our Scott-Knott divided the data if {\em both}
    bootstrapping and an effect size test agreed that the division was statistically significant (99\% confidence) and not a ``small'' effect ($A12 \ge 0.6$).

\section{Results}
\label{sect:results}

{\bf RQ1: Are the default ``off-the-shelf'' parameters for SMOTE appropriate for all
 data sets?}
 
 As discussed above, the default parameters for
 {\sma}, $k,\ m$ and $r$ are $5,\ 50\%$ and $2$.
  Figure~\ref{fig:para} shows the range of parameters
 found by {\smb} across  nine data sets for the 25 repeats of our cross-validation procedure.
 All the results in this figure are {\em within-measure assessment} results, i.e.,
 here, we {\smb}  on a particular performance measure and then we only collect performance for that performance measure on the test set.

 In  Figure~\ref{fig:para}, the {\em median} is the 50th percentile
 value and {\em IQR} is the (75-25)th percentile
 (variance).
 As can be seen in Figure~\ref{fig:para}, most of the learned parameters are far from the default values: 1) Median $k$ is never less than 11; 2) Median $m$ differs according to each data set and quite far from the actual; 3) The $r$ used in the distance function was never 2, rather, it was usually 3.
 Hence,  our answer to {\bf RQ1} is ``no'': the use of off-the-shelf {\sma} should be deprecated. 
 
 We note that many of the settings in Figure~\ref{fig:para} are very similar; for e.g., median values of $k=13$ and $r=3$ seems to be a common
result irrespective of data imbalance percentage among the datasets.  Nevertheless, we do {\em not} recommend replacing
the defaults of {\sma} with the findings
of Figure~\ref{fig:para}. Also, IQR bars are
very large. Clearly, {\smb}'s decisions vary dramatically
depending on what data  is being processed.  Hence,
we strongly recommend that {\smb} be applied to each new data set.

\begin{figure*}[!t]
\begin{minipage}{.49\linewidth}
\centering
        \includegraphics[width=\linewidth ]{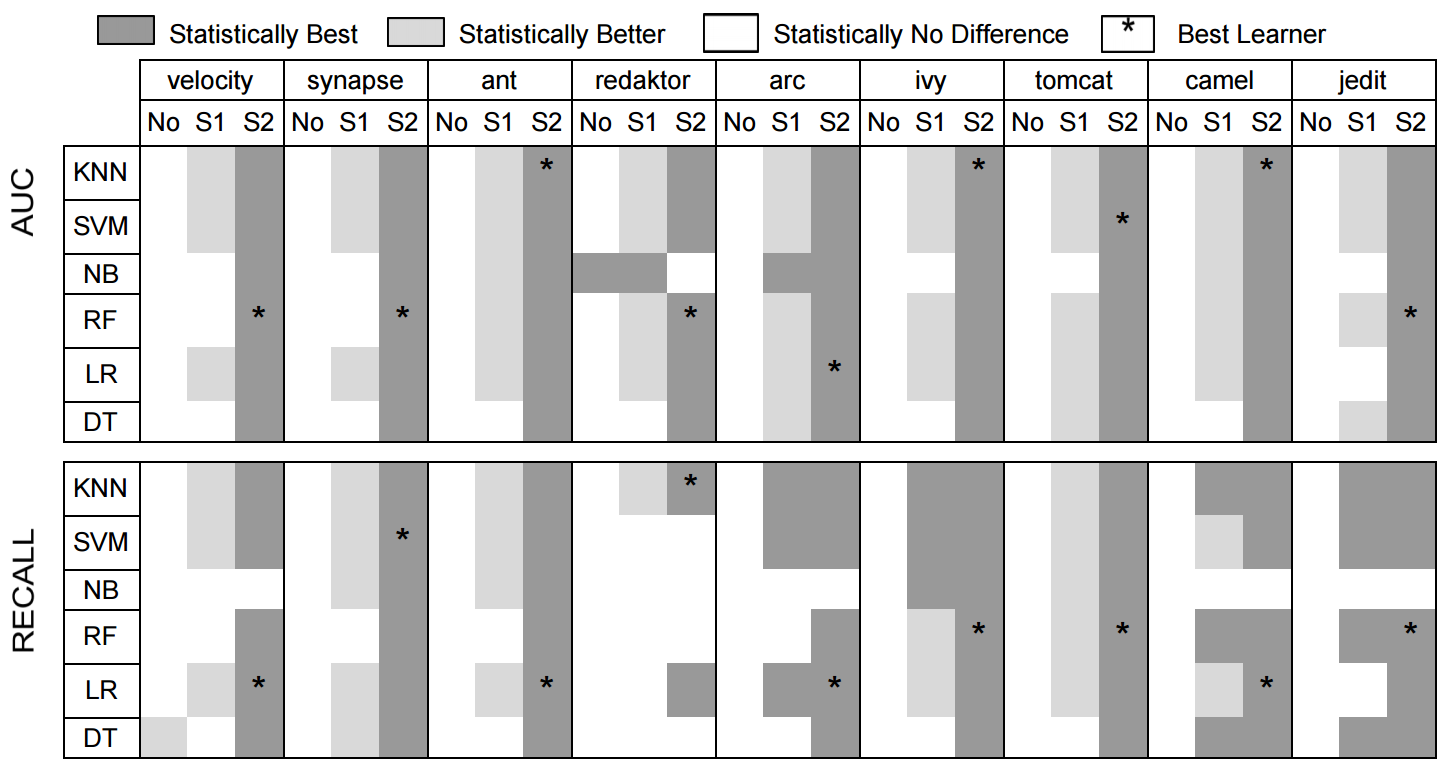}
            \end{minipage}%
\begin{minipage}{.49\linewidth}
        \centering
        \includegraphics[width=\linewidth]{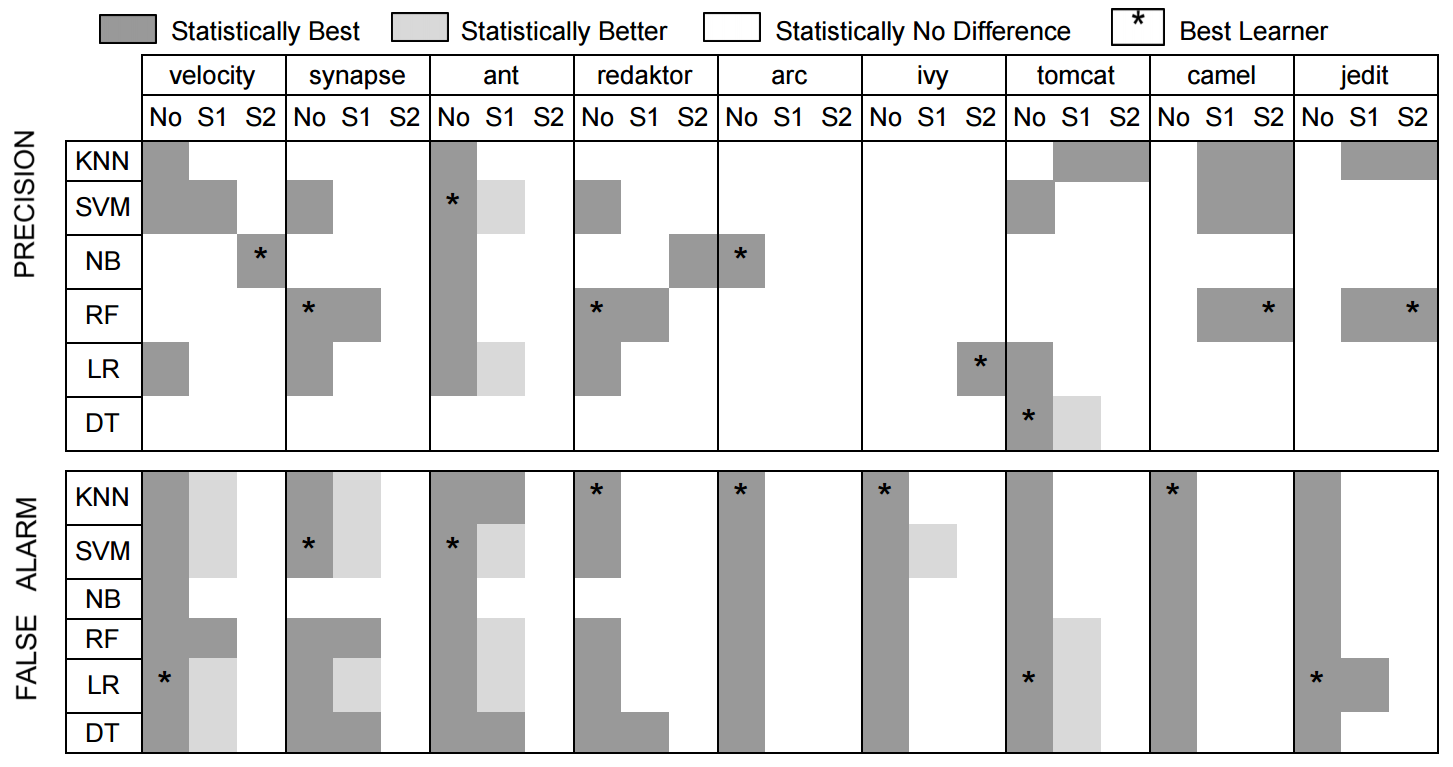}
    \end{minipage}%
    \caption{Scott Knott analysis of No-SMOTE, SMOTE and SMOTUNED. The column headers are denoted as No for No-SMOTE, S1 for SMOTE and S2 for SMOTUNED. $(\ast)$ Mark represents the best learner combined with its techniques.}
    \label{fig:stats}
\vspace{-0.4cm}
\end{figure*}

\noindent
{\bf \\RQ2: Is there any benefit in tuning the default parameters of SMOTE for each new data set?}

Figure~\ref{fig:tuned} shows the performance delta of the {\em within-measure assessment rig}.
Please recall that when this rig applies {\smb}, it optimizes for performance measure, $M_i \in \{recall,\ precision,\ false$ $\ alarm,\ AUC\}$
after which it uses the {\em same} performance measure
$M_i$ when evaluating the test data. In Figure~\ref{fig:tuned}, each subfigure shows that DE is optimized for each $M\_i$ and results are reported against the same $M\_i$.
From the figure~\ref{fig:tuned}, it is observed that
{\smb} achieves large AUC (about 60\%) and recall (about 20\%) improvements relatively
without
 damaging precision and  with only minimal changes
 to false alarm. Another key observation here that can be made is that improvements in AUC with {\smb} is constant whether imbalance is of 34\% or 2\%. Another note should be taken of the AUC improvements, that these are the largest improvements
 we have yet seen, for any prior treatment of defect prediction data. Also, for the raw AUC values, please see \url{http://tiny.cc/raw_auc}.

Figure~\ref{fig:stats} offers a statistical analysis
of different results achieved
after applying our three data pre-filtering methods: 1) {\em \textbf{NO}} = do nothing, 2) {\em \textbf{S1}} = use default {\sma}, and 3) {\em \textbf{S2}} = use {\smb}.
For any learner, there are three such treatments and {\em darker} the cell, {\em better} the performance. 
In that figure, cells with the same color are
either not statistically significantly different or
are different only via a {\em small effect}
(as judged by the statistical methods described in Section~\ref{sec:scott-knott}).

As to what combination of pre-filter+learner works better for any data set, that is marked by a `*'. Since we have three pre-filtering methods and six learners providing us with in-total 18 treatments, and `*' represents the best learner picked with highest median value.

In the  AUC and recall results,  the best ``*'' cell always appears in the S2={\smb} column, i.e.,  
 {\smb} is always
used by the best combination of pre-filter+learner .

As to precision  results,  at first glance, the  results in Figure~\ref{fig:stats} look bad for {\smb} since, less than half the times, 
the best ``*''  happens  in S2={\smb} column.
 But recall from Figure~\ref{fig:tuned} that the absolute size of the precision deltas is very small.  Hence, even though {\smb} ``losses'' in this statistical analysis, the pragmatic impact of that result  is  negligible. But if we can get feedback from domain/expert, we can change between {\sma} and {\smb} dynamically based on the measures and data miners.
 
As to the false alarm results from Figure~\ref{fig:stats}, as discussed above in \tion{performance}, the cost of increased recall is to also increase
the false alarm rate. For e.g., the greatest \textit{increase} in the recall was 0.58 seen in the {\em jEdit} results. This increase comes at a cost
of \textit{increasing} the false alarm rate by 0.20. Apart from this one large outlier, the overall pattern is that the recall improvements range from +0.18 to +0.42 (median to max)
and these come at the cost of much smaller false alarm \textit{increase} of 0.07 to 0.16 (median to max). 
 
In summary, the answer to {\bf RQ2} is that our  AUC and recall results strongly endorse the  use of {\smb}
while the precision and false alarm rates
show there is little harm in using {\smb}.

Before moving to the next research question, we note that these
 results offer an interesting insight on prior ranking studies.
Based on the Ghotra et al.
results of  Table~\ref{tbl:learners}, our expectation was that Random Forests (RF)
would yield the best results across this defect data. Figure~\ref{fig:stats} reports
that, as predicted by Ghotra et al., RF earns more ``stars'' than any other learner, i.e., it is seen to be ``best'' more often than anything else. That said,  
RF was only ``best'' 
in   11/36 of those results, i.e., even our ``best'' learner (RF) fails over half the time.

It is significant to note that
  {\smb} was  consistently  used  by  whatever  learner  was  found  to  be ``best'' (in recall and AUC). 
Hence, we conclude   prior ranking study results (that only assessed different learners) have missed a much more general effect; i.e.  
it can be more useful to reflect on data pre-processors than algorithm selection.
To say that another way, at least for defect prediction,
``better data'' might be better than ``better data miners''.

\noindent
{\bf \\RQ3: In  terms  of  runtimes,  is  the  cost  of  running  SMOTUNED worth the performance improvement?}

\begin{figure}[!htbp]
  \centering
\includegraphics[width=3.3in,keepaspectratio]{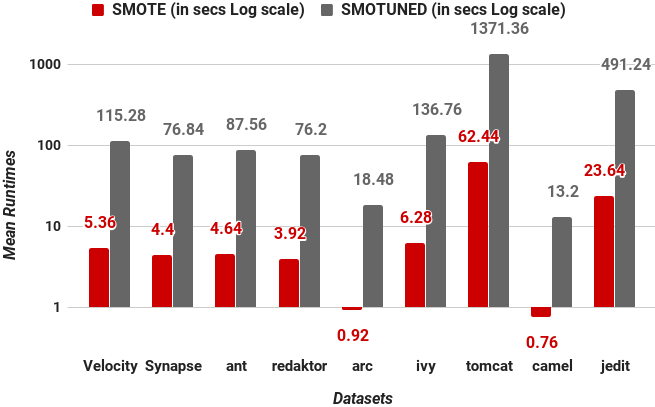}
\vspace{-0.3cm}
  \caption{Data sets vs Runtimes. Note that the numbers
  shown here are the mean times seen across 25 repeats of a 5*5 cross-validation study.
  }
  \label{runtime}
  \vspace{-0.4cm}
\end{figure}

Figure \ref{runtime} shows the mean runtimes
for running a 5*5 cross-validation study for six learners for each data set.
These runtimes were collected from one machine running CENTOS7, with 16 cores.
Note that they do not increase monotonically with the size of the data sets--  a result we can explain with respect to the internal
structure of the data.
Our version of {\sma} uses ball trees to optimize the nearest neighbor calculations. Hence, the runtime of that algorithm is dominated by the internal topology of the data sets rather than the number of classes.
Also, as shown in 
Figure~\ref{fig:pseudocode},
{\smb} explores the local space until it finds $k$ neighbors of the same class. This can take a variable amount of time to terminate.

\begin{figure*}[!t]
\begin{minipage}{.5\linewidth}
\centering
        \includegraphics[width=0.75\linewidth,keepaspectratio,trim=1cm 1cm 1cm 1cm]{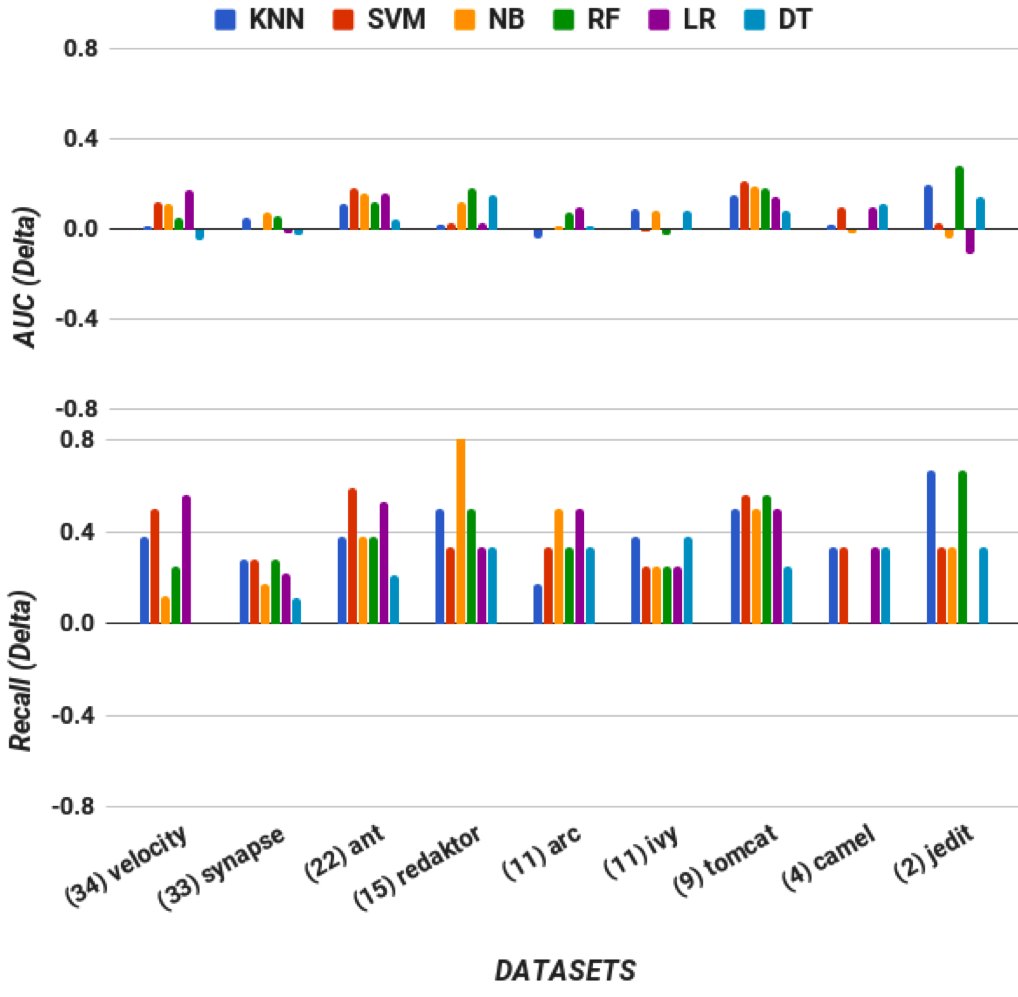}
    \end{minipage}%
\begin{minipage}{.5\linewidth}
        \centering
        \includegraphics[width=0.75\linewidth,keepaspectratio,trim=1cm 1cm 1cm 1cm]{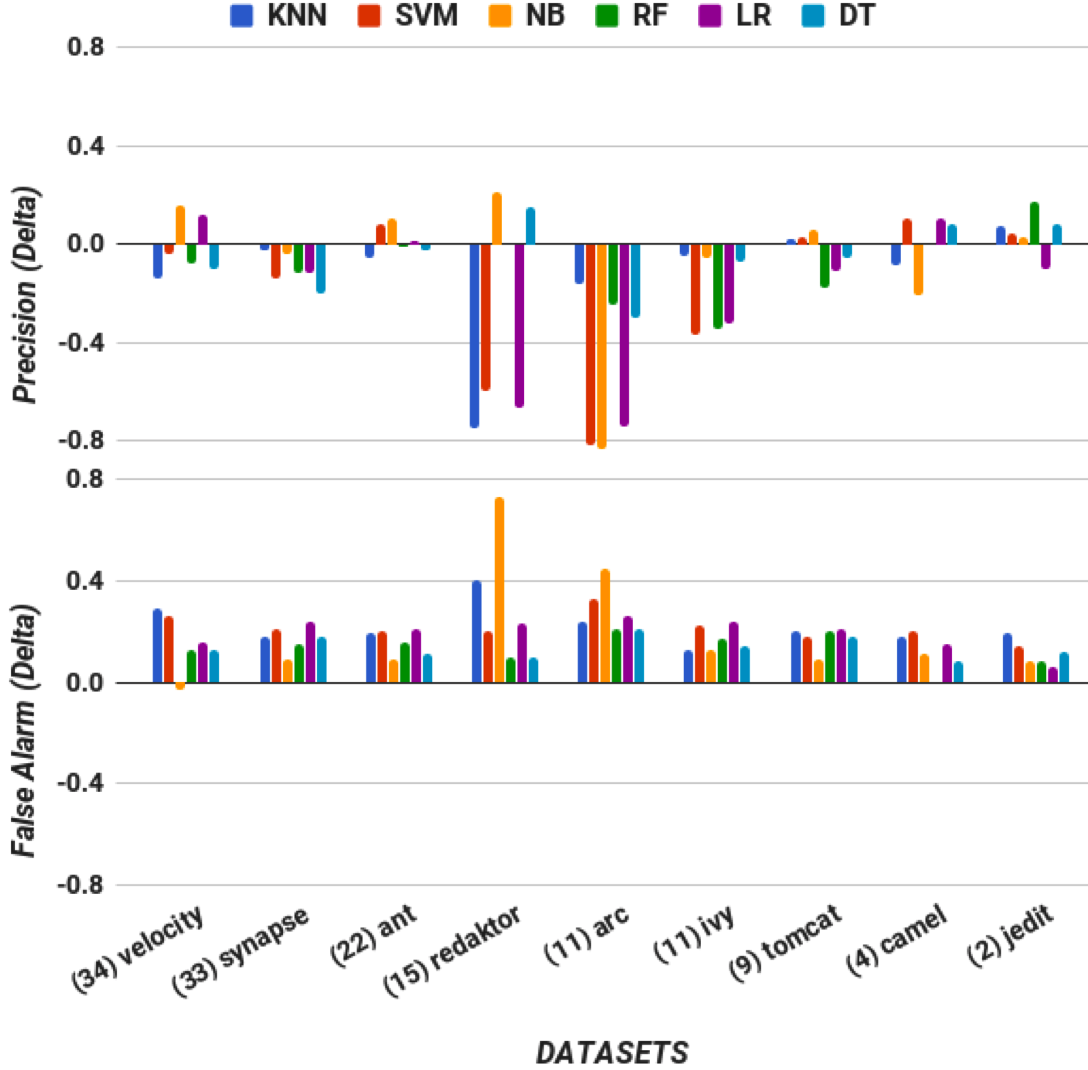}
    \end{minipage}%
    
    \caption{SMOTUNED improvements over MAHAKIL~\cite{bennin2017mahakil}. \underline{Within}-Measure
    assessment (i.e., for each of these charts,
    optimize for performance measure $M_i$, then test for
    performance measure $M_i$). Same format as  Figure~\ref{fig:tuned}.  }
    \label{fig:s2_mahakil}
    \vspace{-0.3cm}
\end{figure*}

As expected,  {\smb} is an order of magnitude slower than {\sma} since
it has to run {\sma} many times to assess different parameter settings.
That said, those runtimes are not excessively slow.
{\smb} usually terminates in under two minutes and never more than half an hour.
Hence, in  our opinion, we answer {\bf RQ3} as ``yes'' since the   performance increment
seen in Figure~\ref{fig:tuned} is more than to compensate for the extra CPU required for {\smb}.

\noindent
{\bf \\RQ4: How does SMOTUNED perform against more recent class imbalance technique?}

All the above work is based on tuning the original 2002 {\sma} paper~\cite{chawla2002smote}. While that
version of {\sma} is widely used in the SE literature, 
it is prudent to compare {\smb} with more recent work.
Our reading
of the literature is that the MAHAKIL algorithm  of Bennin et al.~\cite{bennin2017mahakil} represents the most recent work in  SE on handling  
class imbalance.  
At the time of writing of this paper (early August 2017), there was no reproduction package available for MAHAKIL so we wrote our own version
based on the description in that paper (Available on \url{http://tiny.cc/mahakil}). We verified our implementation on their datasets, and achieved close to their values $\pm$ 0.1. The difference could be due to different random seed. 

Figure~\ref{fig:s2_mahakil} compares   results from MAHAKIL with those from {\smb}. These results
were generated using the same experimental methods as used for   Figure~\ref{fig:tuned} (those methods were described in  Section~\ref{sect:wcm}).
The following table repeats   the statistical analysis of Figure \ref{fig:stats} to report how often
  {\sma}, {\smb}, or MAHAKIL achieves best results across nine data sets.   Note that, in this following table, {\em larger} values are {\em better}:
  
\vspace{-0.4 cm}
{\small\[\begin{array}{*{20}{c|c|c|c|c}}
  & \multicolumn{4}{c}{\text{number of wins}}\\ 
    {\text{Treatments}} & {\text{AUC}} & {\text{Recall}} & {\text{Precision}}& {\text{False Alarm}}\\
    \hline
   {\text{MAHAKIL}} & 1/9 & 0/9 & \textbf{6/9} & \textbf{9/9}  \\
   {\text{SMOTE}}& 0/9 & 1/9 & 0/9 & 0/9 \\
   {\text{SMOTUNED}}& \textbf{8/9} & \textbf{8/9} & 3/9  & 0/9  \\
 \end{array} \]}
\vspace{-0.2 cm}

These statistical tests tell us that the differences seen in Figure~\ref{fig:s2_mahakil}  are large enough to be significant. Looking at Figure~\ref{fig:s2_mahakil}, there are 9 datasets on x-axis, and the differences in precision
  are   so small
in 7 out of those 9 data sets that the pragmatic impact of those differences is small. As to AUC and recall,  we see that 
{\smb} generated   larger and better results than MAHAKIL (especially for recall). {\smb} generates slightly larger false alarms but, in 7/9 data sets,
the increase in the false alarm rate is very small.

According to its authors~\cite{bennin2017mahakil},
MAHAKIL was developed to reduce the false alarm rates on SMOTE and on that criteria it succeeds
 (as seen in  Figure~\ref{fig:s2_mahakil}, since {\smb} does lead to slightly higher false
alarm rates).   But, as discussed above in section \ref{sect:performance}, the downside on minimizing false alarms is also minimizing our ability to 
find defects which is measured in terms of AUC and recall, {\smb} does best.
  Hence,  if this paper was a comparative assessment of  {\smb} vs MAHAKIL, we would conclude that
  by recommending {\smb}.
  
\begin{figure*}[!t]
\begin{minipage}{.5\linewidth}
\centering
        \includegraphics[width=.75\linewidth,keepaspectratio,trim=1cm 1cm 1cm 0cm]{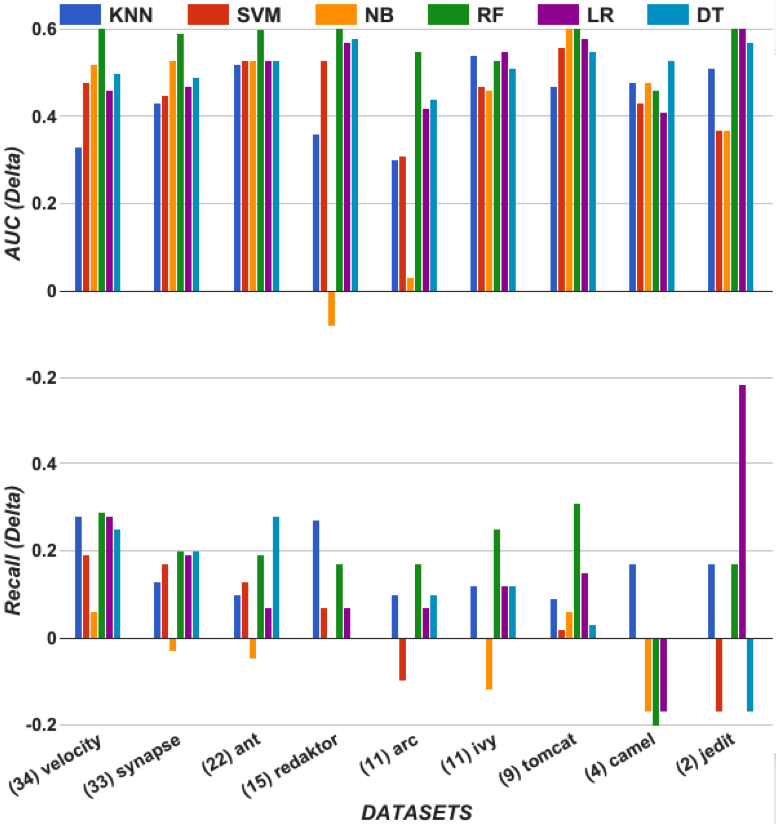}
    \end{minipage}%
\begin{minipage}{.5\linewidth}
        \centering
        \includegraphics[width=.75\linewidth,keepaspectratio,trim=1cm 1cm 1cm 0cm]{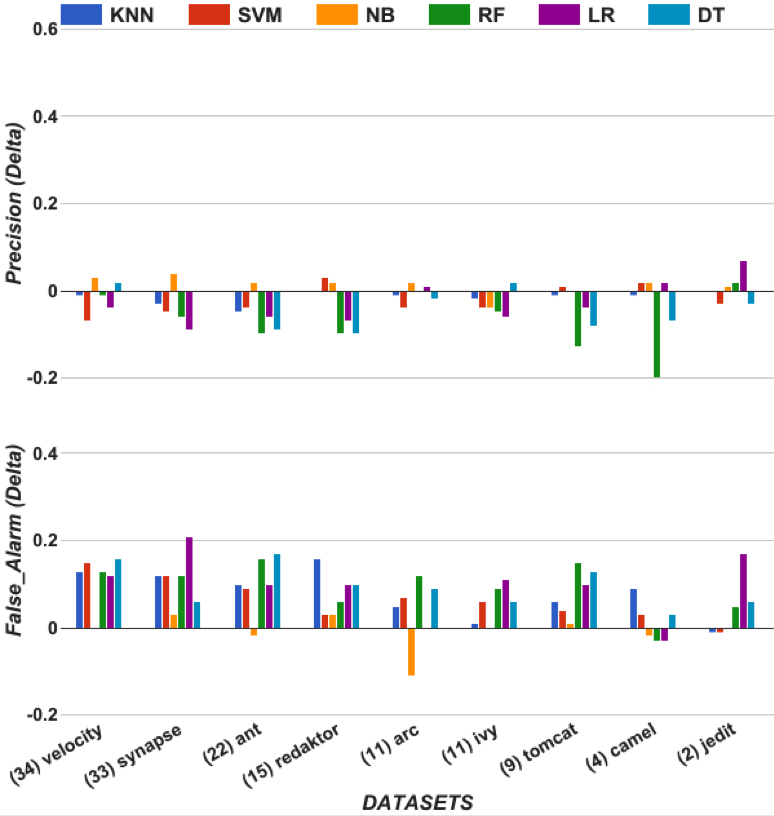}
    \end{minipage}%
    
    \caption{ SMOTUNED improvements over SMOTE. 
    \underline{{\bf Cross}}-Measure
    assessment (i.e., for each of these charts,
    optimize for \underline{{\bf AUC}}, then test for
    performance measure $M_i$).  Same format as
    Figure~\ref{fig:tuned}.}
    \label{fig:auc22}
    \vspace{-0.3cm}
\end{figure*}

However, the goal of this paper is to defend the claim that ``better data'' could be better than ``better
data miners'', i.e., data pre-processing is more effective  than switching to another data miner.
In this regard,   there is something insightful to conclude if we combine the results of {\em both}
MAHAKIL and {\smb}.  
In the MAHAKIL experiments, the researchers spent some time on tuning the learner's parameters. That is,  Figure~\ref{fig:s2_mahakil} is really a comparison
of two treatments: tuned data miners+adjust data against just using {\smb} to adjust the data.
Note that {\smb} still achieves better results  even though the MAHAKIL treatment {\em adjusted both
data and data miners}. Since {\smb} performed so well without tuning the data miners,
we can conclude from the conjunction of these experiments that  ``better data'' is better than using   ``better
data miners''.

Of course, there needs to be further studies done in other SE applications to make the above claim. There is also one more treatment {\em not} discussed in the paper: tuning {\em both}
the data pre-processor {\em and} the data miners. This is a very, very large search space
so while we have experiments running to explore this task, at this time we have not definitive
conclusions to report.

\section{Threats to Validity}
\label{sect:validity}

As with any empirical study, biases can affect the final
results. Therefore, any conclusions made from this work must consider the following issues in mind.

\textbf{\textit{Order bias}}: With each data set how data samples are distributed in training and testing set is completely random. Though there could be times when all good samples are binned into training and testing set. To mitigate this order bias, we run
the experiment 25 times by randomly changing the order of the data samples each time.

\textbf{\textit{Sampling bias}} threatens any classification experiment, i.e., what matters there may not be true here. For e.g., the data sets used here comes from the SEACRAFT repository and were supplied by one individual. These data sets have used in various case studies by various researchers~\cite{he2012investigation,peters2013better,peters2013balancing,turhan2013empirical}, i.e., our results are not more biased than many other studies in this arena.
That said, our nine open-source data sets   are mostly from Apache. Hence
it is an open issue if our results hold for
 proprietary projects and open source projects from other sources.


\textbf{\textit{Evaluation bias}}: In terms of evaluation bias,
our study is far less biased than many other ranking studies.  As shown by our sample of
22 ranking studies in
Table~\ref{tbl:survey2}, 19/22 of those prior studies used {\em fewer} evaluation criteria
than the four reported here (AUC, recall, precision and false alarm). 

The analysis done in RQ4 could be affected by some other settings which we might not have considered since the reproduction package was not available from the original paper~\cite{bennin2017mahakil}.
That said, there is another more subtle evaluation bias arises in the Figure~\ref{fig:tuned}. The four plots of that figure are four {\em different} runs of our  {\em within-measure assessment rig}
(defined in \tion{wcm}). Hence, it is reasonable to check what happens when (a)~one
evaluation criteria is used to control {\smb}, and (b)~the results are assessed
using all four evaluation criteria. 
Figure~\ref{fig:auc22} shows the results of such a {\em cross-measure assessment rig} where AUC was used to control {\smb}. We note that the results in this figure are very similar to Figure~\ref{fig:tuned}, e.g., the precision deltas aver usually tiny, and false alarm increases are usually smaller than the associated recall improvements. But there are some larger improvements in Figure~\ref{fig:tuned}
than Figure~\ref{fig:auc22}. Hence, we recommend cross-measure assessment only if CPU is critically restricted. Otherwise, we think {\smb} should be controlled by whatever is the downstream evaluation criteria
(as done in the within-measure assessment rig of Figure~\ref{fig:tuned}.)

\section{Conclusion}
\label{sect:conclusion}

Prior work on ranking studies tried to improve software analytics by selecting better learners.
Our results show that there may be {\em more} benefits in exploring data pre-processors like {\smb} because we found  that no  learner  was  usually  
``best''  
across all  data  sets  and  all  evaluation  criteria. On one hand, across the same data sets,
{\smb} was  consistently  used  by  whatever  learner  was  found  to  be ``best'' in the  AUC/recall results. On the other hand, for the precision and false alarm results, there was little evidence against the use of {\smb}. That is, creating better training data  (using techniques like {\smb}) may be  more important than  the  subsequent  choice  of a classifier. To say that another way, at least for defect prediction, ``better data'' is  better than ``better
data miners''.

As to specific recommendations, we suggest that any prior ranking study  which did not  study the effects of data pre-processing needs to be analyzed again. Any future such ranking study should include a {\sma}-like
 pre-processor. {\sma} should not be used with its default parameters.
 For each new data set, {\sma} should be used with some automatic parameter tuning tool in
order to find the best parameters for that data set. {\smb} is one of the examples of parameter tuning. Ideally, {\smb} should be tuned using the evaluation criteria used to assess the final predictors. However, if there is not enough CPU to run {\smb} for each new evaluation criteria, {\smb} can be tuned using AUC.



\balance

\bibliographystyle{ACM-Reference-Format}

\end{document}